\documentclass[12pt]{article}
\usepackage{amscd,amsmath,amssymb,amstext,amsthm,exscale,latexsym}
\usepackage{graphicx}
\usepackage{upgreek,txfonts}
\newcommand {\al}   {\alpha}       \newcommand {\bt}  {\beta}
\newcommand {\g }   {\gamma}       
\newcommand {\dl}   {\delta}       \newcommand {\e }  {\epsilon}

\newcommand {\lm}   {\lambda}

\newcommand {\vf }  {\varphi}      
         
\newcommand {\Lm}   {\Lambda}      \newcommand {\Om}  {\Omega}
       
\newcommand {\pl}   {\partial}

\renewcommand {\ln}{{\sf\,ln}}         

\newcommand {\vol}  {\sqrt{{\scriptstyle |}g{\scriptstyle |}}}
            
   \newcommand {\MR}  {{\mathbb R}}

\begin{document}
\title     {Was there a Big Bang?}
\author    {D. E. Afanasev
            \thanks{E-mail: daniilaf@gbu.edu.cn, daniel\_afanasev@yahoo.com}\\
            \sl School of Sciences, Great Bay University, No.16 Daxue Road,\\
            Songshan Lake District, Dongguan 523000, PR China
            \\ M. O. Katanaev
            \thanks{E-mail: katanaev@mi-ras.ru}\\
            \sl Steklov mathematical institute,
            \sl ul.~Gubkina, 8, Moscow, 119991, Russia}
\date      {}
\maketitle
\begin{abstract}
New one parameter family of exact solutions in General Relativity with a scalar
field is found. The metric is of Liouville type which admits complete separation
of variables in the geodesic Hamilton--Jacobi equation. This solution exists for
the exponential potential for a scalar field and is invariant with respect to
global Lorentz transformations. It describes, in particular, evolution of the space-time with the naked singularity. Solutions corresponding to the naked
singularity provide accelerating expansion of the homogeneous and isotropic
Universe, and can be smoothly continued along geodesics to infinite past
without Big Bang.
\end{abstract}
{\em Introduction.}
Exact solutions of Einstein's equations play a crucial role in General
Relativity, providing physical interpretation of the model and comparison
with the observational data. Though many exact solutions are known (see e.g\
\cite{StKrMaHoHe03,GriPod09}), this field of research remains one of the most
wanted and interesting.

We consider General Relativity minimally coupled to a scalar field with
arbitrary potential. Two assumptions are made: (i) the metric is of
Liouville type, i.e.\ it is conformally Lorentzian, the conformal factor being
the sum of arbitrary functions depending on single coordinates, and (ii) the
scalar field depends on coordinates only through the conformal factor. A general
solution of the respective field equations depends on one parameter and is very
simple. It is invariant under global Lorentz transformations and, therefore,
admits six noncommuting Killing vector fields. Note, that symmetry of the
space-time was not assumed from the very beginning but arises as the consequence
of the field equations (spontaneous symmetry emergence
\cite{AfaKat19,AfaKat20A}). The resulting solution is a global one, because any
geodesic can be either continued to infinite value of the canonical parameter or
it ends up at a singularity at its finite value. The obtained metrics describe,
in particular, evolution of the space-time with the naked singularity.

The solutions exist only for the special exponential type of the scalar field
potential. Such potentials attract much interest at present.
They arise in higher-dimensional gravity models, superstring and M-theory (see,
e.g. \cite{BurBar88,Townse03}) and are used in cosmological models
\cite{Halliw87,AnCaKa11,Chakra17}.

Solutions with the naked singularity in the Friedmann form describe the
accelerating expansion of the homogeneous and isotropic Universe only inside the light cone,
which is smoothly continued along geodesics to infinite future and past.

{\em Notation and solution.}
We consider four-dimensional space-time with coordinates $x=(x^\al)$,
$\al=0,1,2,3$. Let there be the Liouville metric
\begin{equation}                                                  \label{aodmkj}
 g_{\al\bt}:=\Phi(x)\eta_{\al\bt},\qquad\eta_{\al\bt}:=\text{diag}(+1,-1,-1,-1),
\end{equation}
where the conformal factor is the sum of four arbitrary functions on single
arguments
\begin{equation}                                                  \label{ajdhtg}
  \Phi:=\phi_0(x^0)+\phi_1(x^1)+\phi_2(x^2)+\phi_3(x^3).
\end{equation}
This is the famous Liouville metric \cite{Liouvi49} well known in mechanics,
because it admits complete separation of variables in the Hamilton--Jacobi
equation for geodesic lines, even if arbitrary functions are not specified.
Note, that it has no Killing vectors in general. It means that the geodesic
equations are Liouville integrable, even if there is no symmetry at all.

The Liouville metric (\ref{ajdhtg}) is conformally flat, its Weyl tensor
vanishes, and, therefore, its curvature tensor is of type {\bf 0} in Petrov's
classification.

The curvature and Ricci tensors and scalar curvature for metric (\ref{aodmkj})
are
\begin{align}                                                    \label{ansmahh}
  R_{\al\bt\g}{}^\dl:=&\pl_\al\Gamma_{\bt\g}{}^\dl-\pl_\bt\Gamma_{\al\g}{}^\dl
  -\Gamma_{\al\g}{}^\e\Gamma_{\bt\e}{}^\dl+\Gamma_{\bt\g}{}^\e
  \Gamma_{\al\e}{}^\dl=
\\                                                                     \nonumber
  =&\frac1{2\Phi}\big(\Phi''_{\al\g}\dl_\bt^\dl-\Phi''_{\bt\g}\dl_\al^\dl
  -\Phi''_\al{}^\dl\eta_{\bt\g}+\Phi''_\bt{}^\dl\eta_{\al\g}\big)+
\\                                                                     \nonumber
  &+\frac3{4\Phi^2}\big(-\phi'_\al\phi'_\g\dl_\bt^\dl+\phi'_\bt\phi'_\g\dl_\al^\dl
  +\phi'_\al\phi^{\prime\dl}\eta_{\bt\g}-\phi'_\bt\phi^{\prime\dl}\eta_{\al\g}
  \big)+
\\                                                                     \nonumber
  &+\frac1{4\Phi^2}\phi'_\e\phi^{\prime\e}\big(\eta_{\al\g}\dl_\bt^\dl
  -\eta_{\bt\g}\dl_\al^\dl\big),
\\                                                                \label{eassxv}
  R_{\al\g}:=&R_{\al\bt\g}{}^\bt=
  \frac1{2\Phi}\big(2\Phi''_{\al\g}+\Phi''_\e{}^\e\eta_{\al\g}\big)
  -\frac3{2\Phi^2}\phi'_\al\phi'_\g,
\\                                                                \label{egsjsd}
  R:=&g^{\al\g}R_{\al\g}=\frac3{\Phi^2}\Phi''_\al{}^\al-\frac3{2\Phi^3}\phi'_\al
  \phi^{\prime\al},
\end{align}
where $\Gamma_{\al\bt}{}^\g$ are Christoffel's symbols.
Here and in what follows raising and lowering of indices are performed by using
the Lorentz metric $\eta_{\al\bt}$, and the prime denotes derivatives with
respect to the corresponding arguments.

If $\Phi>0$, then the signature of the metric is $(+---)$, and
a scalar field $\vf(x)$ in General Relativity with cosmological constant $\Lm$
is described by the action
\begin{equation}                                                  \label{ancsgt}
  S:=\int\!dx\vol\left[R-2\Lm+\frac12g^{\al\bt}\pl_\al\vf\pl_\bt\vf-V(\vf)
  \right],\qquad g:=\text{det} g_{\al\bt},
\end{equation}
where the potential for a scalar field $V(\vf)$ will be specified later. Now we
only assume that it is bounded from below.

Variation of action (\ref{ancsgt}) with respect to all metric components
$g_{\al\bt}$ and a scalar field yields field equations. For the Liouville metric
(\ref{aodmkj}), they are
\begin{align}                                                     \label{abnaas}
  R_{\al\bt}=-\frac12\pl_\al\vf\pl_\bt\vf+\frac12\Phi\eta_{\al\bt}
  (V+2V),
\\                                                                \label{aqsfge}
  \eta^{\al\bt}\pl^2_{\al\bt}\vf+\frac1\Phi\phi^{\prime\al}\pl_\al\vf+\Phi V'=0,
\end{align}
which are to be solved.

We shall look for solutions of the field equations assuming that the scalar
field depends on coordinates through the conformal factor,
$\vf(x):=\Psi\big(\Phi(x)\big)$, where $\Psi\big(\Phi\big)$ is an arbitrary
function of single argument. Then equation (\ref{abnaas}) for $\al\ne\bt$
implies
\begin{equation*}
  \frac{3\phi'_\al\phi'_\bt}{\Phi^2}=\Psi^{\prime 2}\phi'_\al\phi'_\bt,
\end{equation*}
where $\Psi':=d\Psi/d\Phi$. Therefore, function $\Psi$ satisfies equation
\begin{equation*}
  \Psi^{\prime 2}=\frac3{\Phi^2}.
\end{equation*}
For nonconstant functions $\phi_\alpha$ its general solution is
\begin{equation}                                                  \label{afoiuy}
  \vf:=\Psi=\pm\sqrt3\ln\big(C\Phi\big)\qquad\Leftrightarrow\qquad
  C\Phi=\text{e}^{\pm\vf/\sqrt3},
\end{equation}
where $C>0$ is an arbitrary integration constant. Then Eqs.~(\ref{abnaas}),
(\ref{aqsfge}) reduce to
\begin{align}                                                     \label{ahdesj}
  \Phi''_{\al\bt}+\frac12\Phi''_\g{}^\g\eta_{\al\bt}=&\frac12\eta_{\al\bt}\Phi^2
  (V+2\Lm),
\\                                                                \label{awshgf}
  \Phi''_\g{}^\g\pm\frac{\Phi^2}{\sqrt3}V'=&0.
\end{align}
Taking the trace of Eq.~(\ref{ahdesj}) we obtain
\begin{equation}                                                  \label{ajkskh}
  3\Phi''_\g{}^\g=2\Phi^2(V+2\Lm).
\end{equation}
This equation coincides with Eq.~(\ref{awshgf}) if and only if the potential
satisfies the differential equation
\begin{equation}                                                  \label{akshfb}
  V+2\Lm\pm\frac{\sqrt3}2V'=0.
\end{equation}
It has a general solution
\begin{equation*}
  |V+2\Lm|=\text{e}^{\mp\frac2{\sqrt3}(\vf+\vf_0)},
\end{equation*}
where $\vf_0$ is an integration constant. Assuming that $V+2\Lm$ is bounded
from below for all $\vf$, the modulus sign can be dropped. So, the potential is
\begin{equation}                                                  \label{amdfks}
  V=-2\Lm+\text{e}^{\mp\frac2{\sqrt3}(\vf+\vf_0)}.
\end{equation}

Equation (\ref{akshfb}) has also the singular solution $V\equiv-2\Lm$
corresponding to the massless scalar field, which is not considered here.

Now the only equation which has to be solved is Eq.~(\ref{ahdesj}), which takes
the form
\begin{equation}                                                  \label{edkfuy}
  \Phi''_{\al\bt}=\frac16\eta_{\al\bt}\Phi^2(V+2\Lm).
\end{equation}
For the Liouville conformal factor (\ref{ajdhtg}), $\Phi''_{\al\bt}\equiv0$ when
$\al\ne\bt$. Therefore, off diagonal components of this equation are satisfied.
The diagonal components reduce to
\begin{equation*}
\begin{split}
  \phi''_0=&~~\frac16\Phi^2(V+2\Lm),
\\
  \phi''_i=&-\frac16\Phi^2(V+2\Lm),\qquad i=1,2,3.
\end{split}
\end{equation*}
Taking the sum of these equations we get
\begin{equation*}
  \phi''_0+\phi''_i=0,\qquad\forall i.
\end{equation*}
The summands depend on different arguments and hence are equal to the same
positive constant $b_0$ with opposite signs:
\begin{equation}                                                  \label{absjlp}
\begin{split}
  \phi_0=&~~b_0(x^0-x^0_0)^2+c_0,
\\
  \phi_i=&-b_0(x^i-x^i_0)^2+c_i,
\end{split}
\end{equation}
where $x^\al_0$ and $c_\al$ are arbitrary integration constants.
Shifting the coordinates we put $x^\al_0=0$. Moreover, rescaling the coordinates
$x^\al\mapsto kx^\al$, where $k:=b_0^{-1/4}$, constant $b_0$ can be set to
unity. Hence the conformal factor is the quadratic polynomial:
\begin{equation}                                                  \label{avskls}
  \Phi=s+c,
\end{equation}
where
\begin{equation*}
  s:=\eta_{\al\bt}x^\al x^\bt, \qquad c:=c_0+c_1+c_2+c_3\in\MR.
\end{equation*}
Now Eq.~(\ref{edkfuy}) reduces to
\begin{equation}                                                  \label{ajfhyt}
  2=\frac16\Phi^2(V+2\Lm),
\end{equation}
and Eq.~(\ref{awshgf}) becomes
\begin{equation}                                                  \label{adjskl}
  2=\frac1{6C^2}\text{e}^{\mp\frac2{\sqrt 3}\vf_0},
\end{equation}
where Eqs.~(\ref{afoiuy}) and (\ref{amdfks}) were used. It defines constant
$\vf_0$ in terms of $C$. Therefore, the potential (\ref{amdfks}) becomes
\begin{equation}                                                  \label{amdfps}
  V=-2\Lm+12C^2\text{e}^{\mp\frac2{\sqrt3}\vf}.
\end{equation}

Thus, the particular solution of field Eqs.~(\ref{abnaas}) and
(\ref{aqsfge}) with ansatz $\vf:=\Psi(\Phi)$ is found for metric signature
$(+---)$. The solution is very simple: the conformal factor in the Liouville
metric and the scalar field are given by Eqs.~(\ref{avskls}) and (\ref{afoiuy}).

For negative conformal factor, $\Phi<0$, the metric signature is $(-+++)$ and the model with the action (\ref{ancsgt}) implies appearance of a ghost. We consider this case as a toy model of the black hole formation. Note, that the solution with the equivalent metric exists for the signature $(+---)$ in the case of the potential (\ref{amdfps}) with minus sign, that is unbounded from below. For convenience, we keep the potential bounded from below and consider two opposite signatures. Therefore, the obtained metric is given by
Eqs.~(\ref{aodmkj}) and (\ref{avskls}) with $\Phi>0$ and $\Phi<0$ for signatures
$(+---)$ and $(-+++)$, respectively.

Scalar field (\ref{afoiuy}) diverges as $s\to-c$ or $s\to\pm\infty$. It is
constant on two sheeted, $s=\text{const}>0$, and one sheeted,
$s=\text{const}<0$, hyperboloids and on two cones $s=0$.

So, we have obtained one parameter family of metrics
\begin{equation}                                                  \label{ansdhg}
  g_{\al\bt}=\Phi\eta_{\al\bt}=(s+c)\eta_{\al\bt},
\end{equation}
which are defined in regions $s+c>0$ and $s+c<0$ for signatures $(+---)$ and
$(-+++)$, respectively.
This metric is obviously invariant with respect to global Lorentz rotations.
Therefore, it has six noncommuting Killing vector fields. Note, that we did not
assume any symmetry of the metric at the very beginning. It appears due to the
field equations. This phenomena was called spontaneous symmetry emergence in
\cite{AfaKat19,AfaKat20A}. Sure, metric (\ref{ansdhg}) is also spherically
symmetric, because the rotational group is the subgroup of the Lorentz group.

The Hamilton--Jacobi equation for geodesics for metric (\ref{ansdhg}) admits
complete separation of variables and belongs to class $[0,4,0]_2$ according to
the classification given in \cite{Katana23A,Katana23B}. The geodesic Hamiltonian
equations in this coordinate system have four independent involutive quadratic
conservation laws.

It is interesting that metric (\ref{ansdhg}) nontrivially depends on time $x^0$
and spacial coordinates, and there is no coordinate system where it is static
even locally for $s>0$.

Curvature tensor (\ref{ansmahh}) for the obtained solution is
\begin{equation}                                                  \label{asndbv}
\begin{split}
  R_{\al\bt\g}{}^\dl=&\frac1\Phi\left(3-\frac c\Phi\right)
  \big(\eta_{\al\g}\dl_\bt^\dl-\eta_{\bt\g}\dl_\al^\dl)-
\\
  &-\frac3{\Phi^2}\big(x_\al x_\g\dl_\bt^\dl-x_\bt x_\g\dl_\al^\dl
  -x_\al x^\dl\eta_{\bt\g}+x_\bt x^\dl\eta_{\al\g}\big).
\end{split}
\end{equation}
It is asymptotically flat for $s\to\pm\infty$, but we cannot say that the space-time is
asymptotically Minkowskian, because metric becomes degenerate there. 

The simplest curvature invariants are
\begin{align}                                                     \label{amdndh}
  R^{\al\bt\g\dl}R_{\al\bt\g\dl}=&\frac{12}{\Phi^4}\left(9-6\frac c\Phi
  +5\frac{c^2}{\Phi^2}\right),
\\                                                                \label{ajghdk}
  R^{\al\bt}R_{\al\bt}=&\frac{36}{\Phi^4}\left(3+\frac{c^2}{\Phi^2}\right),
\\                                                                \label{edgffd}
  R=&\frac{6}{\Phi^2}\left(3+\frac c\Phi\right),
\end{align}
We see, that curvature is singular for $\Phi=0$, the singularity being located on
one-sheeted, two-sheeted hyperboloids or cone, depending on constant $c$.

It can be proved that the space-time is maximally extended along geodesics, i.e.\
any geodesic line can be either extended to infinite value of the canonical
parameter or it ends up at a singularity at its finite value. Therefore,
coordinates $(x^\al)$ are the global ones.

\begin{figure}[hbt]
\hfill\includegraphics[width=.9\textwidth]{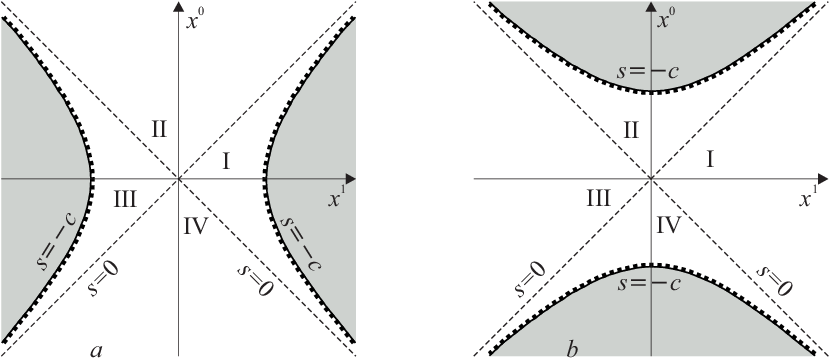}
\hfill {}
\centering\caption{The Liouville metric for $\Phi>0$, $c>0$ ({\em a}) and
$\Phi<0$, $c<0$ ({\em b}).}
\label{Liouville}
\end{figure}

We draw the allowed regions of coordinates on the $x^0,x^1$ slice for $\Phi>0$
and $c>0$  (the cases $c\le0$ will be considered elsewhere) in
Fig.~\ref{Liouville}{\em a}. Note, that this picture has to be rotated in
two extra space dimensions around the $x^0$ axis. Therefore, the forbidden
region (grey) is the connected one-sheeted ``hyperboloid''. Its boundary is the
time-like naked singularity. Test particles can either move throw the throat of
the singularity and live forever or fall on the singularity at a finite proper
time.

In Fig.~\ref{Liouville}{\em b}, the allowed region of coordinates is shown for
$\Phi<0$ and $c<0$. Test particles can live forever in the Universe (the rotated
quadrants I or III) or fall into the black hole in quadrant II at a finite
proper time, the cone $s=0$ being the horizon. This space-time describes
formation of the black hole. The space-time in the far past is represented by the
rotated quadrants I or III. There is no singularity there for $x^0<0$ except the
white hole in quadrant IV. At a finite proper time corresponding to $s=0$ the
horizon appears, and afterwards the space-like singularity is formed.

If null geodesic starts at any point inside the horizon in quadrant II (Fig. 1b)
and goes ahead in time, then it inevitably hits singularity at a finite value of
the canonical parameter. The same happens also for all time-like geodesics,
because the horizon is the cone $s=0$. This is the black hole in the usual
sense: no signal can go out from inside of the horizon (at the classical level)
and everything falls into the singularity for a finite proper time. For a fixed
moment of time $x^0$, the horizon is a sphere of finite radius, it appears with
infinitesimal radius at $x^0=0$, and goes to infinity during time evolution. We
see, that the horizon is dynamical. This is in great difference with the
Schwarzschild horizon, which is a two dimensional sphere of constant radius.

The Liouville metric (\ref{ansdhg}) can be rewritten in the Friedmann form by
introduction of pseudospherical coordinates in quadrant II:
\begin{equation*}
\begin{split}
  x^0:=& t\cosh\chi,
\\
  x^1:=& t\sinh\chi\sin\theta\cos\varphiup,
\\
  x^2:=& t\sinh\chi\sin\theta\sin\varphiup,
\\
  x^3:=& t\sinh\chi\cos\theta,
\end{split}
\end{equation*}
where
\begin{equation*}
   t>0,\qquad\chi\in(0,\infty),\qquad\theta\in(0,\pi),\qquad
  \varphiup\in(0,2\pi).
\end{equation*}
Then the metric is
\begin{equation}                                                  \label{agmtju}
  ds^2=(t^2+c)(dt^2-t^2d\Om),
\end{equation}
where
\begin{equation*}
  d\Om:=d\chi^2+\sinh^2\chi(d\theta^2+\sin^2\theta d\varphiup^2)
\end{equation*}
is the metric on the north sheet of two sheeted hyperboloid embedded in
Minkowskian space-time $\MR^{1,3}$ (constant negative curvature three dimensional
Riemannian manifold). These coordinates cover quadrant II in
Fig.~\ref{Liouville}{\em a,b} rotated in two extra space dimensions. The
rotated quadrant IV is covered by the same coordinates with replacement
$t\mapsto-t$.

If $\Phi>0$, we introduce new time coordinate $\tau$ defined by the differential
equation
\begin{equation*}
  \frac{d\tau}{d t}=\sqrt{ t^2+c},\qquad t^2+c>0
\end{equation*}
Then
\begin{equation}                                                  \label{anbdgf}
  \tau-\tau_0=\frac{ t\sqrt{ t^2+c}}2+\frac c2\ln\left|
   t+\sqrt{ t^2+c}\right|,\qquad\tau_0=\text{const}.
\end{equation}
The respective metric becomes
\begin{equation*}
  ds^2=d\tau^2-a(\tau)^2d\Om,
\end{equation*}
and the scale factor
\begin{equation*}
   a(\tau)= t\sqrt{ t^2+c}
\end{equation*}
is defined implicitly by Eq.~(\ref{anbdgf}). Its derivatives are
\begin{equation}                                                  \label{abansg}
\begin{split}
  \frac{da}{d\tau}=&\frac{2 t^2+c}{ t^2+c},
\\
  \frac{d^2a}{d\tau^2}=&\frac{2c t}{( t^2+c)^{5/2}}.
\end{split}
\end{equation}
We see, that the Universe in quadrant II is expanding with acceleration, constant
velocity and deceleration for $c>0$, $c=0$, and $c<0$, respectively. In quadrant
IV, the situation is opposite. Thus, expansion with acceleration takes place
for the naked singularity corresponding to $c>0$ in quadrant II. This expansion
starts from the ``horizon'', $t=0$, with zero scale factor, and there is no
singularity in global coordinates. Going back in time, the homogeneity and
isotropy of space sections are lost after crossing the ``horizon'' at finite
proper time, and an observer turns out in the throat of the naked singularity.
He sees signals from naked singularity and a hole in the center. Time-like
worldlines can go through this hole and be extended to infinity. There is no Big
Bang in this scenario.

For $\Phi<0$ and $c:=-\lm^2<0$, $\lm>0$ (otherwise the solution is not defined
in quadrant II), the new coordinate $\tau$ is given by equation
\begin{equation*}
  \frac{d\tau}{dt}=\sqrt{\lm^2-t^2},\qquad \lm^2-t^2>0.
\end{equation*}
Its general solution is
\begin{equation*}
  \tau-\tau_0=\frac{t\sqrt{\lm^2-t^2}}2
  +\frac{\lm^2}2\arcsin\left(\frac t\lm\right).
\end{equation*}
Now the scale factor
\begin{equation*}
  a(\tau)=t\sqrt{\lm^2-t^2}
\end{equation*}
has derivatives:
\begin{equation*}
\begin{split}
  \frac{da}{d\tau}=&~~\frac{\lm^2-2 t^2}{\lm^2-t^2},
\\
  \frac{d^2a}{d\tau^2}=&-\frac{2\lm^2 t}{(\lm^2-t^2)^{5/2}}.
\end{split}
\end{equation*}
The Universe expands for $0<t<\!\lm/\sqrt2$ and contracts for
$\lm/\!\sqrt2<t<\lm$ with deceleration. The scale factor is zero both on the
horizon and black hole. In quadrant IV, the Universe expands for
$-\lm<t<-\lm/\sqrt2$ and contracts for $-\lm/\sqrt2<t<0$ with deceleration.

In quadrants I or III the solution cannot be brought to the Friedmann form,
because hypersurfaces $s=\text{const}<0$ are time-like.

{\em Conclusion.}
The problem of a black hole formation has a long history started from the
seminal paper \cite{OppSny39}. Oppenheimer and Snyder assumed, in particular,
that everything is spherically symmetric, the energy-momentum tensor of a star
is produced by fluid-like matter, and the metric is of the Schwarzschild form
outside the star. These assumptions were used in many subsequent papers. The
essential point here is that the energy-momentum tensor is not obtained from the
variation of matter action with respect to metric components and, in addition,
there may be a problem in gluing smoothly solutions inside and outside a star.
There were another approaches to matter collapse obtaining matter
energy-momentum tensor from the variational principle (see, e.g.\
\cite{Chakra17} and references therein). These models, to our knowledge, were
solved either approximately or with strong simplifying assumptions different
from ours.

In the present paper, a new simple exact global solution of Einstein equations
with a scalar field is found. The solution has several novel features:
1) it is invariant under global Lorentz transformations;
2) it depends nontrivially both on time and space coordinates;
3) there are four quadratic independent involutive conservation laws for
geodesics in global coordinates $x^\al$;
4) it describes, in particular, travelling through the naked singularity and
the black hole formation. As far as we know, it is essentially different from
all previously known solutions. For $\Phi>0$ and $c>0$, solutions
describe accelerating expansion of the homogeneous and isotropic Universe
under ``horizon'', the scale factor there being zero. After crossing the
``horizon'' back in time, space sections loose their
homogeneity and isotropy. This region corresponds to the throat of the naked
singularity. An observer sees light rays from naked singularity, and there is
the hole. Time-like worldlines can be extended through it to infinity. The
obtained solutions explicitly show that cosmological solutions in General
Relativity can be smoothly extended through the zero of the scale factor.
This property raises the question used as the title of the paper. In addition,
there is no reason to add dark energy to the model, its role is playing by the
scalar field with reasonable physical properties. Hopefully, the obtained
Liouville solution will help us in deeper understanding of General Relativity,
black hole formation and cosmology.

{\em Acknowledgement.}
The work of M.O. Katanaev was performed at the Steklov
International Mathematical Center and supported by the Ministry of Science and
Higher Education of the Russian Federation (agreement no. 075-15-2022-265).


\begin{thebibliography}{10}

\bibitem{StKrMaHoHe03}
H.~Stephani, D.~Kramer, M.~A.~H. MacCallum, C.~Hoenselaers, and E.~Hertl.
\newblock {\em Exact Solutions of Einstein's Field Equations}.
\newblock Cambridge University Press, Cambridge, 2003.

\bibitem{GriPod09}
J.~B. Griffiths and Podolsk\'y.
\newblock {\em Exact Space-times in Einstein's General Relativity}.
\newblock Cambridge University Press, Cambridge, 2009.

\bibitem{AfaKat19}
Afanasev~D. E. and M.~O. Katanaev.
\newblock Global properties of warped solutions in general relativity with an
  electromagnetic field and a cosmological constant.
\newblock {\em Phys.\ Rev.\ D}, 100(2):024052, 2019.
\newblock https://doi.org/10.1103/PhysRevD.100.024052
  http://arxiv.org/abs/arXiv:1904.04648 [physics.gen-ph].

\bibitem{AfaKat20A}
Afanasev~D. E. and M.~O. Katanaev.
\newblock Global properties of warped solutions in general relativity with an
  electromagnetic field and a cosmological constant. {II}.
\newblock {\em Phys.\ Rev.\ D}, 101(12):124025, 2020.
\newblock https://doi.org/10.1103/PhysRevD.101.124025.
  http://arxiv.org/abs/2006.09209 [gr-qc].

\bibitem{BurBar88}
A.~B. Burd and J.~D. Barrow.
\newblock Inflationary models with exponential potentials.
\newblock {\em Nucl.\ Phys.}, B308:929--945, 1989.

\bibitem{Townse03}
P.~K. Townsend.
\newblock Cosmic acceleration and {M}-theory.
\newblock arXiv:hep-th/0308149v2 29 Aug 2003.

\bibitem{Halliw87}
J.~J. Halliwell.
\newblock Scalar fields in cosmology with an exponental potential.
\newblock {\em Phys.\ Lett.}, B185(3-4):341--344, 1987.

\bibitem{AnCaKa11}
A.~A. Andrianov, F.~Cannata, and A.~Yu. Kamenshchik.
\newblock General solution of scalar field cosmology with a (piecewise)
  exponential potential.
\newblock {\em JCAP}, 10(004):19 pp., 2011.

\bibitem{Chakra17}
S.~Chakrabarti.
\newblock Scalar field collapce with an exponential potential.
\newblock {\em Gen.\ Rel.\ Grav.}, 49:24(2):1--12, 2017.

\bibitem{Liouvi49}
L.~Liouville.
\newblock M\'emoire sur l'int\'egration des \'equations diff\'erentielles du
  mouvement d'un nombre quelconque de points mat\'eriels.
\newblock {\em J.\ Math.\ Pures Appl.}, 14:257--299, 1849.

\bibitem{Katana23A}
M.~O. Katanaev.
\newblock Complete separation of variables in the geodesic {H}amilton-{J}acobi
  equation.
\newblock {\em arXiv:2305.02222 [gr-qc]}, 2023.

\bibitem{Katana23B}
M.~O. Katanaev.
\newblock Complete separation of variables in the geodesic {H}amilton-{J}acobi
  equation in four dimensions.
\newblock {\em Phys.\ Scr.}, 98(10):104001, 2023.
\newblock DOI 10.1088/1402-4896/acf251.

\bibitem{OppSny39}
J.~R. Oppenheimer and H.~Snyder.
\newblock On continued gravitationsl contraction.
\newblock {\em Phys.\ Rev.}, 56:455--459, 1939.

\end{thebibliography}

\end{document}